\documentclass[aps,prl,twocolumn,showpacs,superscriptaddress]{revtex4-1}  % for review and %\documentclass[aps,prl,twocolumn,showpacs,superscriptaddress,groupedaddress]{revtex4-1}  % for review and submission
\usepackage{graphicx}  % needed for figures
\usepackage{dcolumn}   % needed for some tables
\usepackage{bm}        % for math
\usepackage{amssymb}   % for math
\usepackage{subfigure}
\usepackage{color,epsfig,rotate}
\begin{document}

\title{Field-induced gapless electron pocket in the superconducting
vortex phase of YNi$_2$B$_2$C as probed by magnetoacoustic quantum
oscillations$^\dagger$}
%\thanks{This work is dedicated to the memory of Lev Gor'kov.}
\author{J.\ N\"ossler}
\affiliation{Hochfeld-Magnetlabor Dresden (HLD-EMFL), Helmholtz-Zentrum
Dresden-Rossendorf, 01314 Dresden, Germany}
\affiliation{Institut f\"ur Festk\"orperphysik, TU Dresden,
01062 Dresden, Germany}
\author{R.\ Seerig}
\affiliation{Hochfeld-Magnetlabor Dresden (HLD-EMFL), Helmholtz-Zentrum
Dresden-Rossendorf, 01314 Dresden, Germany}
\affiliation{Institut f\"ur Festk\"orperphysik, TU Dresden,
01062 Dresden, Germany}
\author{S.\ Yasin}
\affiliation{Hochfeld-Magnetlabor Dresden (HLD-EMFL), Helmholtz-Zentrum
Dresden-Rossendorf, 01314 Dresden, Germany}
\author{M.\ Uhlarz}
\affiliation{Hochfeld-Magnetlabor Dresden (HLD-EMFL), Helmholtz-Zentrum
Dresden-Rossendorf, 01314 Dresden, Germany}
\author{S.\ Zherlitsyn}
\affiliation{Hochfeld-Magnetlabor Dresden (HLD-EMFL), Helmholtz-Zentrum
Dresden-Rossendorf, 01314 Dresden, Germany}
\author{G.~Behr}
\thanks{Deceased}
\affiliation{Leibniz Institute for Solid State and Materials
Research, IFW Dresden, 01171 Dresden, Germany}
\author{ S.-L.\ Drechsler}
\affiliation{Leibniz Institute for Solid State and Materials
Research, IFW Dresden, 01171 Dresden, Germany}
\author{G.\ Fuchs}
\affiliation{Leibniz Institute for Solid State and Materials
Research, IFW Dresden, 01171 Dresden, Germany}
\author{ H.\ Rosner}
\affiliation{Max Planck Institute for Chemical Physics of Solids,
01187 Dresden, Germany}
\author{J.\ Wosnitza}
\affiliation{Hochfeld-Magnetlabor Dresden (HLD-EMFL), Helmholtz-Zentrum
Dresden-Rossendorf, 01314 Dresden, Germany}
\affiliation{Institut f\"ur Festk\"orperphysik, TU Dresden,
01062 Dresden, Germany}
\date{\today}

\begin{abstract}
By use of ultrasound studies we resolved magneto-acoustic quantum
oscillation deep into the mixed state of the multiband nonmagnetic
superconductor YNi$_2$B$_2$C. Below the upper critical field, only
a very weak additional damping appears that can be well explained
by the field inhomogeneity caused by the flux-line lattice in the
mixed state. This is clear evidence for no or a vanishingly small
gap for one of the bands, namely, the spheroidal $\alpha$ band. This
contrasts de Haas--van Alphen data obtained by use of torque
magnetometry for the same sample, with a rapidly vanishing oscillation
signal in the mixed state. This points to a strongly distorted
flux-line lattice in the latter case that, in general, can hamper a
reliable extraction of gap parameters by use of such techniques.\\
\\
$^\dagger$This work is dedicated to the memory of Lev Petrovich Gor'kov

\end{abstract}

%\pacs{71.18.+y, 74.70.Dd, 74.25.Jb, 74.25.Uv}
\maketitle

The observation of magnetic quantum oscillations is usually taken as
evidence for the existence of a Fermi surface. The appearance
of Landau levels in a magnetic
field leads to an oscillating density of states at the Fermi
level as a function of field. Experimentally, these oscillations can
be detected in many thermodynamic and transport properties, with the
most prominent being the de Haas--van Alphen (dHvA) effect in the
magnetization. Consequently, the observation of dHvA oscillations in
the mixed state of a superconductor first appeared as a surprise
\cite{gra76}. Below the upper critical field, $B_{c2}$, the opening
of a superconducting gap and the corresponding disappearance of the
entire Fermi surface seem to contradict the existence of such oscillations.
Nevertheless, they were observed in many type-II superconductors
(\cite{jan98,man01,ber12} and references therein).

Motivated by the experimental evidence, however, it subsequently was
shown by a number of theoretical studies that this phenomenon may
be understood in principle in a rather general context. Thereby,
different models are used to explain the occurrence of quantum
oscillations below $B_{c2}$ \cite{mak91,mil93,miy93,duk95,yas02},
but it still remains unclear which of them is the most appropriate.
Usually, the validity of these theories is tested by comparing the
predicted additional damping of the quantum oscillations below
$B_{c2}$ with experiment. Such analysis gives as the main fit
parameter the superconducting gap at zero temperature, $\Delta_0$,
a value that largely depends on the used model.

With the proper theory at hand dHvA data could, in principle, yield
information on the field and angular evolution of the superconducting
gap, $\Delta$. However, considerable ambiguity in $\Delta$ is introduced
not only by the various theoretical predictions but even more
from varying, sometimes contradictory, experimental data. In particular,
for YNi$_2$B$_2$C and LuNi$_2$B$_2$C highly controversial results were
reported \cite{ber12,hei95,gol96,gol97,ter95,ter97,bin03,ign05,ber07,iss08}.
Thereby, for the so-called $\alpha$ band, an additional damping of the dHvA
signal was found either in line with the opening of a weak-coupling gap
\cite{gol96,gol97,ter95,bin03}, or yielding an unexpectedly small gap
\cite{ber12,hei95,ter97,ber07,iss08}, or even an abrupt vanishing of
the oscillations below $B_{c2}$ \cite{ign05}. All these experimental
results were obtained by measuring the dHvA effect by use of either
the torque or the field-modulation method.

Here, we present quantum-oscillation data obtained for YNi$_2$B$_2$C
by use of ultrasound measurements comparing them to magnetic-torque data.
By studying the magnetoacoustic quantum oscillations in the normal and
superconducting state we find a marginal additional damping of the
oscillations below $B_{c2}$, that can be ascribed solely to the expected
magnetic-field inhomogeneities originating from a regular flux-line
lattice, evidencing a vanishingly small or even the absence of a gap on
the $\alpha$ band. This contrasts the abrupt vanishing of the torque
dHvA signal measured for the same sample.

Recently, Gor'kov \cite{gor13} argued that small gaps can be
induced from dominant large, strongly interacting Fermi surfaces
onto small, minor bands of multiband superconductors. He explicitly
mentioned YNi$_2$B$_2$C as one candidate among others.
Furthermore, Barzykin \cite{bar09} and Barzykin and Gor'kov \cite{bar07},
based on a simplified analysis of two special BCS-type two-band models,
proposed the possibility of a partial quenching of superconductivity by
an external magnetic field, i.e., a gapless state in weakly coupled bands
at low temperatures for multiband superconductors. Here, we provide further
general arguments for such a scenario, focusing on the case of a small
weakly coupled pocket relevant for YNi$_2$B$_2$C.

Superconductivity in $R$Ni$_2$B$_2$C ($R$ = rare earth) was discovered
more than 20 years ago \cite{nag94,cav94} and immediately received
considerable attention. (For recent reviews see \cite{mul08,maz15}.)
The unusual properties in the superconducting
state led to controversial debates on its nature. In particular, these
materials were among the first for which multiband superconductivity
was detected \cite{shu98,muk05,yok06,ber08}. Furthermore, pronounced
gap anisotropies have been suggested \cite{mana01,kaw16}.
In previous dHvA studies in
the normal state, band- and angular-dependent mass enhancements evidenced
largely varying coupling strengths for and within the various bands of
LuNi$_2$B$_2$C \cite{iss08,ber08}. Angular-resolved
photoemission-spectroscopy data as well revealed highly anisotropic and
band-dependent gaps in YNi$_2$B$_2$C \cite{bab10}.

High-quality YNi$_2$B$_2$C single crystals were grown by a zone-melting
method \cite{beh99,beh00}, improved by optical heating. From the
resulting rod a piece (HKZ066-A) with dimensions $2.64\times 1.63
\times 1.365$ mm$^3$ has been cut out that successively was annealed
at 900$^\circ$C. This sample was then investigated using ultrasound
and magnetic-torque measurements. It has a superconducting
transition temperature $T_c = 15.3(1)$ K with a transition width of
about 0.1~K and a resistance ratio between 300 and 16 K of 39.

\begin{figure}
	\centering
	\includegraphics[width=0.95\columnwidth]{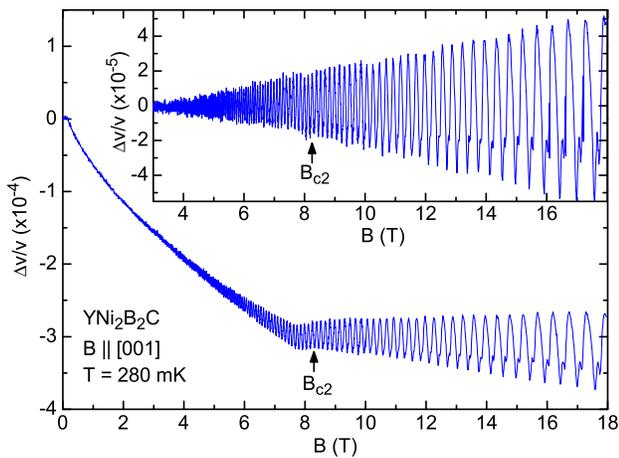}
	\caption{\label{fig:US_dHvA} Magnetic-field dependence
of the relative change of the longitudinal sound velocity for sound
propagating along the $c$ axis in YNi$_2$B$_2$C
measured at 0.28 K in fields up to 18 T aligned as well along $c$.
The inset shows the oscillating part of the signal after background
subtraction.}
\end{figure}

The ultrasound data were taken using a phase-sensitive detection
technique that allowed us to measure the relative changes of the
sound velocity, $\Delta v/v$, and the sound attenuation (not
discussed further here) \cite{lut05}. Thin-film transducers for
longitudinally polarized sound waves propagating at a frequency of
71.3~MHz along the $c$ direction were glued to the polished ends
of the sample. The measurements were done using a $^3$He cryostat
placed inside an 18/20 T superconducting magnet. The dHvA signal
was measured by use of a capacitive cantilever immersed in the $^3$He
of a toploading cryostat in magnetic fields up to 13~T. For that,
the same sample was glued by a small amount of Apiezon-N grease onto
a 50-$\mu$m-thick copper-beryllium cantilever.

The relative change of the sound velocity in YNi$_2$B$_2$C measured
at 280 mK is shown in Fig.\ \ref{fig:US_dHvA}. On a smoothly varying
background signal clear quantum oscillations are visible. With increasing
magnetic field in the superconducting state, the background sound
velocity decreases (lattice softening) until at $B_{c2} \approx 8.25$~T
a small anomaly appears \cite{rem_bc2}, above which the background stays
nearly constant. After subtracting this background the magnetic quantum
oscillations, resolvable above about 3 T, are nicely seen in the inset
of Fig.\ \ref{fig:US_dHvA}. The frequency of this oscillation,
$F_\alpha = 505(1)$ T, agrees well with previous dHvA results
\cite{hei95,gol96,ter95,ter97,bin03,ngu96,rem_harm} and has been ascribed
to a small spheroidal Fermi-surface pocket centered around
the $\Gamma$ point of the Brillouin zone \cite{yam04}. The amplitude
of the oscillating signals smoothly grows with increasing field without
any obvious anomalous changes when going from the mixed to the
normal-conducting state. No hysteresis could be observed between up
and down field sweeps in the ultrasound data.

\begin{figure}
  \centering
  \includegraphics[width=0.95\columnwidth]{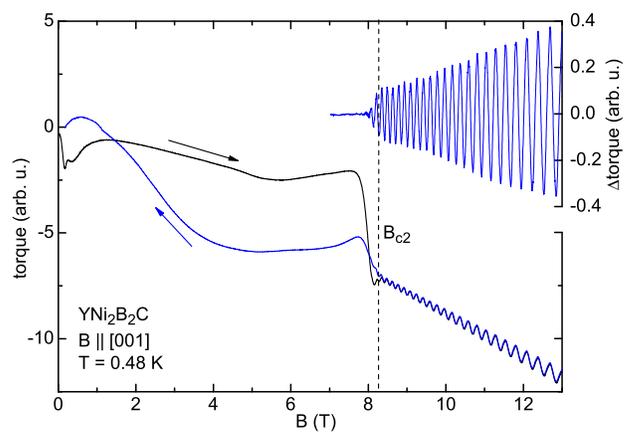}
  \caption{\label{fig:torque} Field dependence of the magnetic
  torque of YNi$_2$B$_2$C. Up and down sweeps are shown (indicated
  by arrows) for fields aligned nearly parallel to $c$.
  In the upper right of the panel the background-subtracted
  signal is shown.}
\end{figure}

This is largely different for the dHvA signal of the same sample in
the torque data (Fig.\ \ref{fig:torque}). In the normal state, the
frequency $F_\alpha$ is well resolvable with smoothly decreasing
amplitude down to $B_{c2}$. In the mixed state, however, the
oscillations abruptly vanish (inset of Fig.\ \ref{fig:torque}).
In the as-measured torque signal a large hysteresis and peak effect
occurs. Such behavior is well known from previous studies
\cite{ter97,ign05,rem_tau}. Nevertheless, the present data allow for
a reliable subtraction of the background signal evidencing the
extremely rapid disappearance of the dHvA signal below $B_{c2}$.

Such a fast vanishing of the dHvA in the mixed state, reported
previously as well in Ref.\ \cite{ign05}, is very unexpected and
cannot be reasonably explained by the opening of a superconducting
gap. The latter is clearly proven by our ultrasound data
for which quantum oscillations persist deep into the mixed
state. On the reason for this strong damping of the torque dHvA
signal we can only speculate at the moment. A possible scenario
is the existence of a strongly disordered vortex arrangement in
the torque experiments. This disorder then leads to pronounced
field inhomogeneities within the sample. Further evidence for
strong disorder in the vortex lattice is given by the large peak
effect appearing in the magnetization, especially close to $B_{c2}$.
This peak effect is connected with a massive rearrangement of
vortices within the sample. Such a feature is expected to be
strongly sample dependent as indeed seen experimentally
\cite{ber12,hei95,gol96,gol97,ter95,ter97,bin03,ign05,ber07,iss08}.
For the ultrasound measurements the persistence of the magnetic
quantum oscillations down to very low fields in the mixed state proves
that in this case no strong field inhomogeneities exist. This might
be caused by the sound waves distorting the crystallographic lattice
and leading thereby to a rearrangement of the vortices into a
regular lattice. In this context, one may refer to earlier
vortex-shaking experiments, where the application of an additional
oscillating magnetic field leads to a fast depinning of the vortex
lattice \cite{wil98,bra02,lin01,mar15}.
Although sound waves do not directly shake vortices, they shake
pinning centers which in return could rearrange the vortices.

Anyway, the direct comparison between torque and ultrasound data
on the same sample evidences that the strong damping in the torque
dHvA signal below $B_{c2}$ cannot be related to an intrinsic opening
of a superconducting gap. Indeed, when analyzing the additional damping
of the dHvA signal using Eq.\ (\ref{maki}), discussed in detail
below, an unphysical huge zero-temperature gap of $\Delta_0 \approx 25$
meV would result. This contrasts the oscillating ultrasound signal
that favorably can be analyzed invoking reasonable damping factors.

\begin{figure}[t]
	\centering
	\includegraphics[width=0.95\columnwidth]{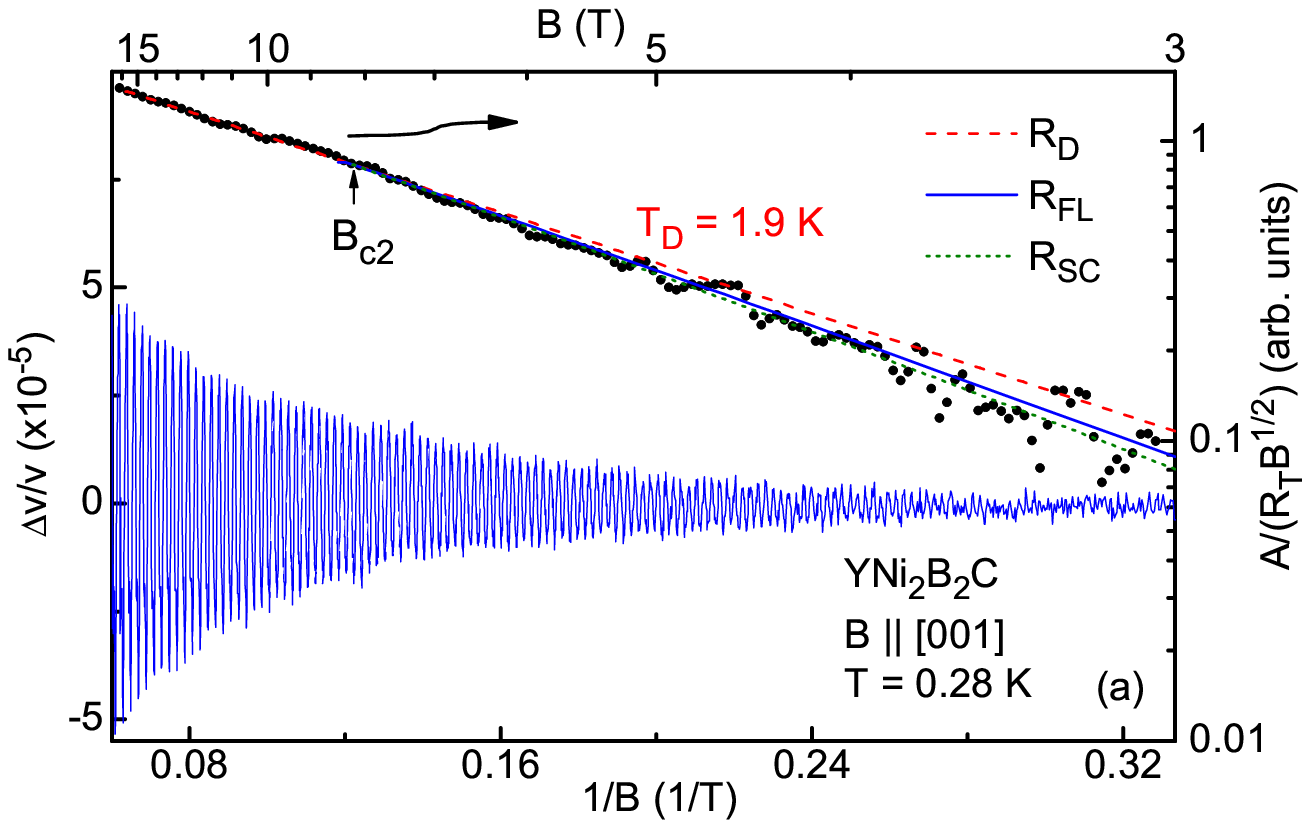}
	\includegraphics[width=0.95\columnwidth]{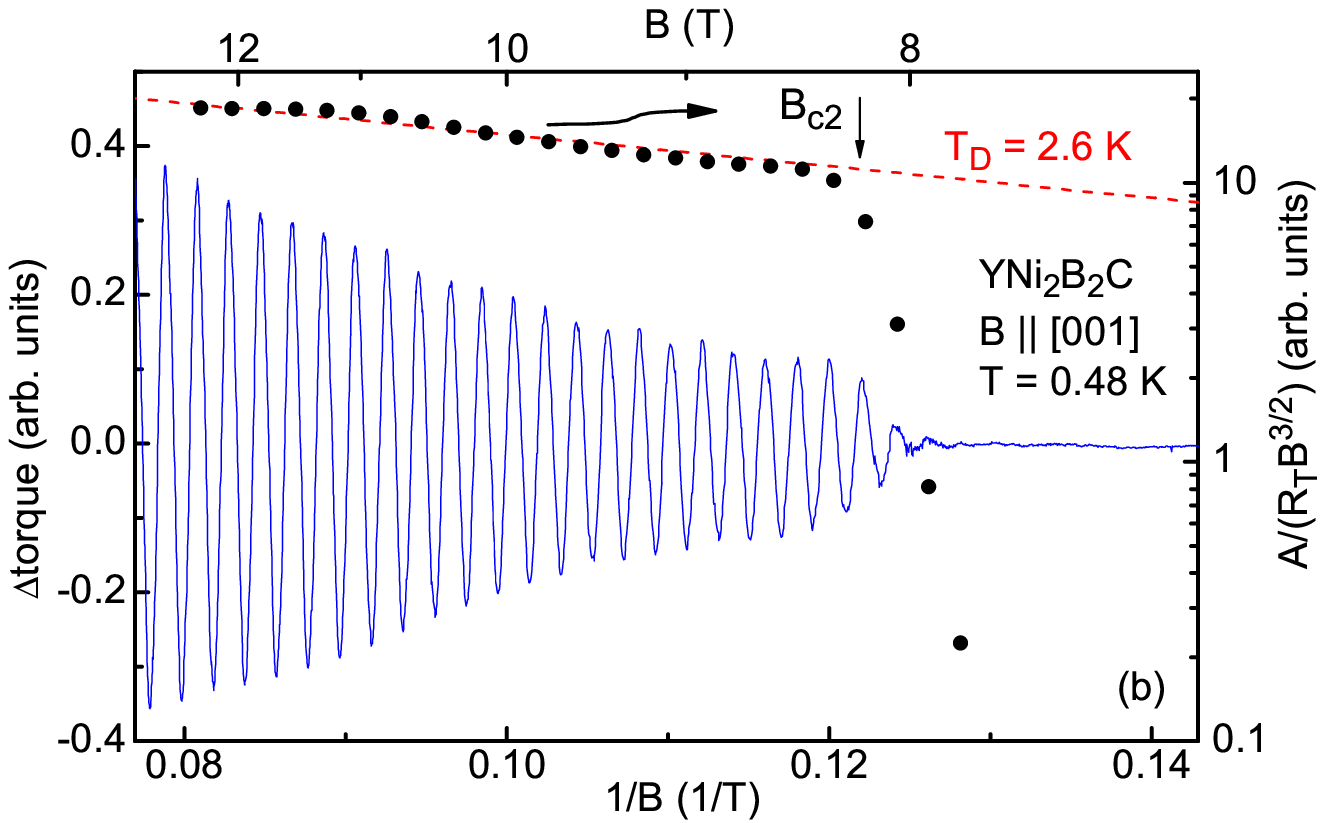}
	\caption{\label{fig:Dingle} (a) Magneto-acoustic
quantum oscillations (see also the inset of Fig.\ \ref{fig:US_dHvA})
together with a Dingle plot of the oscillation amplitude (right axis)
as a function of $1/B$. The dashed line is a fit to the Dingle data
in the normal state. For the solid and dotted lines see main text.
(b) The same kind of plot for the dHvA data measured by use of the
torque method (Fig.\ \ref{fig:torque}) with fit (dashed line)
to the normal-state Dingle data.}
\end{figure}

The field-dependent damping of magnetic quantum oscillations
is usually described by factors \cite{jan98,sho84,rem_fund}
\begin{equation}\label{damp}
    R_i = \exp\left(-\frac{\pi m_c}{eB\tau_i}\right),
\end{equation}
with $e$ the electron charge, $m_c$ the cyclotron effective
mass extracted from the temperature-dependent damping of the quantum
oscillations \cite{rem_mc}, and $\tau_i$ the various ``scattering times''
explained below. We obtain $m_c = 0.34(1)m_e$ for the $\alpha$ orbit,
with $m_e$ the free-electron mass, both from our torque and
ultrasound data. This nicely agrees with previous reports
\cite{hei95,ter95,ter97}. In the mixed state, the effective mass
does not change within error bars as obtained from our ultrasound
data.

In the normal-conducting state, the relevant factor is the Dingle
damping, $R_D$, with the electronic scattering time $\tau_D =
\hbar/(2\pi k_B T_D)$, where $k_B$ is the Boltzmann constant.
The Dingle temperature $T_D$ provides a measure of the scattering
rate within a certain sample and band. When plotting
the amplitudes, $A$, of the magnetic quantum oscillations
with appropriate scalings in a Dingle plot \cite{rem_scal}, $T_D$
is easily extracted from a linear fit to the data above $B_{c2}$
(dashed lines in Fig.\ \ref{fig:Dingle}). We find $T_D = 1.9(1)$ K
for the ultrasound and $T_D = 2.6(4)$ K for the torque data. The
larger error bar for the latter data originates in the limited
available fit range.

In Fig.\ \ref{fig:Dingle}, again the striking difference of the
damping in the mixed state is apparent: In the torque measurement,
the amplitude vanishes abruptly within two or three oscillation
periods [Fig.\ \ref{fig:Dingle}(b)]. Note, that here the
field-dependent amplitudes were determined by Fourier transformation
over three oscillation periods and shifting this window consecutively
by one period. In the ultrasound data [Fig.\ \ref{fig:Dingle}(a)],
only a very small additional damping appears below $B_{c2}$. One
reason for such a damping is the always present field inhomogeneity
caused by the vortex lattice. This can be described by the damping
factor $R_{FL}$ \cite{jan98}, containing the scattering rate
\begin{equation}\label{fll}
    \tau_{FL}^{-1} = \sqrt{\frac{\pi F}{2}}\frac{e}{\pi \kappa^2 m_c}
    \frac{B_{c2} - B}{\sqrt{B}},
\end{equation}
where $\kappa$ is the Ginzburg-Landau parameter and $F$ is the
dHvA frequency ($F_\alpha$ here). Strictly speaking,
this derivation is valid only close to $B_{c2}$ and for an ideal
hexagonal flux-line lattice \cite{rem_hexa}. For YNi$_2$B$_2$C,
$\kappa$ values between 10 and 15 are reported \cite{lee94,hon94}.
Using $\kappa = 12$, we can already well account for the additional
damping seen in the magnetoacoustic quantum oscillations [solid
line in Fig.\ \ref{fig:Dingle}(a)]. Here, we have, however, not yet
considered any damping due to the opening of a superconducting gap.

In fact, in case such a gap would open over the detected $\alpha$
pocket further damping of the oscillating signal would be expected.
All theories considering this gap opening predict a considerable
additional damping \cite{jan98,man01,mak91,mil93,miy93,duk95,yas02}.
Thereby, most of these theories are valid only close to $B_{c2}$.
An often used approach to describe the additional damping in the
superconducting state is that introduced by Maki which results in
the damping term $R_{SC}$ as given by Eq.\ (\ref{damp}) with
scattering rate \cite{mak91}
\begin{equation}\label{maki}
    \tau_{SC}^{-1} = \Delta^2\frac{m_c}{e\hbar B}
    \sqrt{\frac{\pi B}{F}},
\end{equation}
where $\Delta$ is the field-dependent superconducting gap averaged
over the cyclotron orbit. With the usual approximation
$\Delta = \Delta_0 \sqrt{1-B/B_{c2}}$, the only fit parameter is
$\Delta_0$. For the total damping, now consisting of $R_{D}$, $R_{FL}$,
and $R_{SC}$, we obtain a maximum gap of $\Delta_0 \approx 0.6$ meV
\cite{rem_gap}. This total damping is shown by the dotted line in
Fig.\ \ref{fig:Dingle}(a). In the weak-coupling limit, $\Delta_0 =
1.764 k_B T_c \approx 2.33$ meV would be expected, in accord with
the $\alpha$-band-gap value obtained in a recent ab-initio study
\cite{kaw16}. We emphasize here, that the factor $R_{SC}$ is not
necessary to describe the ultrasound data; even with $R_{SC} = 1$
($\Delta_0 = 0$) the fit is excellent.

\begin{figure}
	\centering
	\includegraphics[width=0.48\columnwidth]{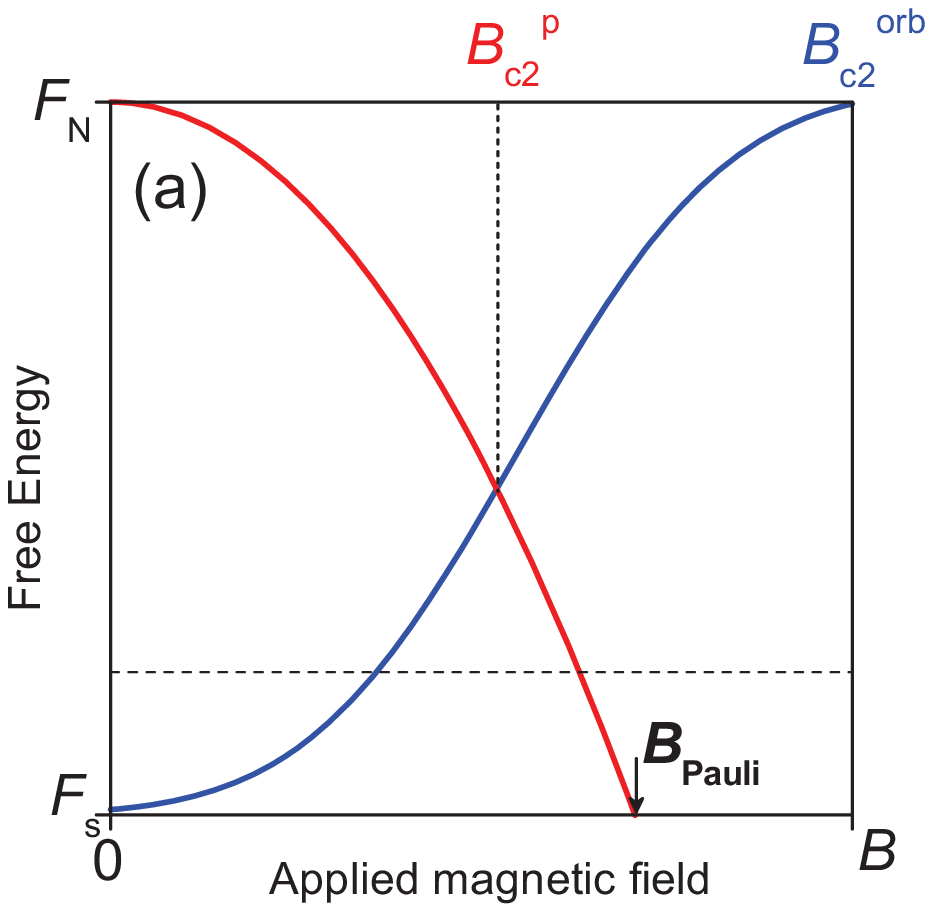}
	\includegraphics[width=0.48\columnwidth]{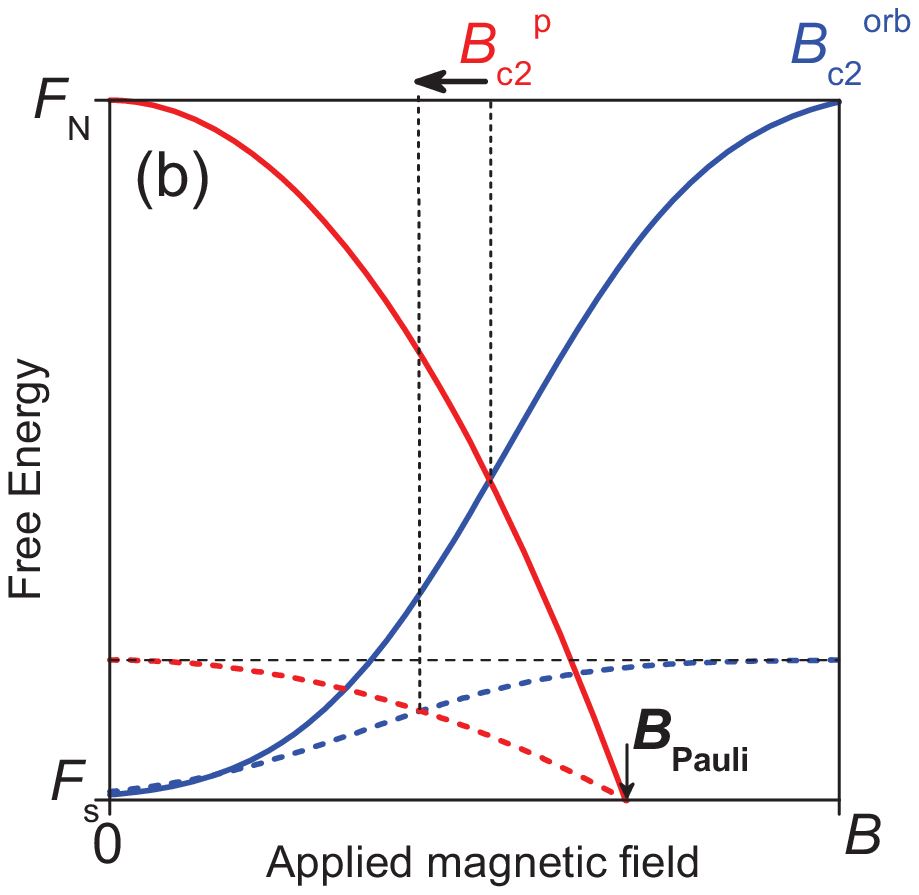}
	\caption{\label{fig:Pauli} (a) Schematic view of the determination
    of $B_{c2}$ from the energy-gain balance between the condensation
    and field penetration for a single-band superconductor, after
    \cite{WHH,fuc09}. (b) The same for an artificially isolated
    weakly coupled band of a multiband superconductor with almost
    vanishing interband coupling to the dominant band which is still
    approximately described as in (a) \cite{rem_plot}.}
\end{figure}

A possible scenario explaining our experimental observation, using
a realistic multiband approach sketched below, is
the field-induced quenching of superconductivity in the $\alpha$
pocket well below $B_{c2}$. As mentioned in the introduction such
behavior may occur in multiband superconductors for a weakly
with otherwise strongly coupled bands \cite{bar09,bar07}.
A schematic sketch elucidating the quenching of superconductivity,
i.e., the reduction of the critical field in a weakly coupled
band is shown in Fig.\ \ref{fig:Pauli}. In order to estimate
such a reduced critical field, $B_{c2}^\alpha(0)$, we employ
a generalized band-specific expression derived within isotropic
Eliashberg theory \cite{shu02}, valid here for $B \parallel c$,
\begin{equation}\label{bc2alpha}
B_{c2,c}^\alpha(0) \approx K \frac{T^2_c(1+\lambda_{\alpha})^{2.4}}
{v^2_{F,\alpha,ab}}
\left(1 + \frac{0.13\gamma_{imp}}{T_c\left(1+
\lambda_{\alpha}\right)}\right),
\end{equation}
where $\lambda_\alpha$ is the effective $\alpha$-band coupling
constant, $v_{F,\alpha,ab}$ is the $\alpha$-band Fermi velocity in the
$ab$ plane, $\gamma_{imp}\approx 2\pi T_D \approx 12$ K is the impurity
scattering rate, and the prefactor $K = k_B^2\pi^2\exp(2-\gamma)/
(2\hbar e) = 2.31\times 10^8$ V/(K$^2$s), with the Euler constant
$\gamma = 0.577$. Using the experimental $T_c = 15.3$ K, the
renormalized Fermi velocity $v_{F,\alpha,ab}/(1 + \lambda_{\alpha})
\approx 4.19\times 10^5$ m/s from dHvA measurements \cite{win01}, and
a reasonable range of $\lambda_{\alpha}$ between 0.51 and 1.35
\cite{all75}, in accord with Ref.\ \cite{kaw16}, rather modest
$B_{c2,c}^\alpha(0)$ values between 0.39 and 0.45 T are obtained.
This estimate implies that superconductivity in the $\alpha$
pocket should be quenched for fields well below 3 T above which
the dHvA oscillations can be resolved.

A more quantitative self-consistent multiband description within this
scenario is outside the scope of the present work and will be considered
elsewhere. Anyhow, our suggested weak interband-coupling clean-limit
scenario provides additional support for the quenching of a small
Fermi-surface pocket in already weak fields. Hence, in the mixed state
the $\alpha$ pocket of YNi$_2$B$_2$C has indeed very likely a vanishingly
small or zero gap while (most of) the other Fermi-surface sheets
develop gaps in the superconducting state. These gaps
have varying, anisotropic, and partially large values for the
different bands as was evidenced by various studies
\cite{shu98,muk05,yok06,hua06}. Further, we suggest that
a similar field-induced quenching mechanism might resolve the
puzzle of the accidental point nodes ($s$+$g$ wave) proposed by
Maki et al. \cite{mak02} based on thermal conductivity \cite{iza02}
and NMR \cite{sai07} studies performed in fields above 1 T.

In conclusion, we found strong experimental evidence, well supported
by theoretical arguments, for the existence of a gapless or, at least,
marginally small gapped band in the mixed state of YNi$_2$B$_2$C.
This is proven by our magnetoacoustic quantum-oscillation data that
persist deep into the mixed state. It also contrasts our torque
dHvA data that would suggest an unphysical large gap. The latter most
likely is caused by a strongly disordered flux-line arrangement in the
sample. Our results emphasize that, in general, great care is needed when
superconducting gap values are being extracted from magnetic quantum
oscillations, especially when the flux-line distribution is strongly
distorted. Finally, our quenching approach is of interest as well
for other multiband superconductors, such as the iron-based materials,
having dominant and minor bands, too.

Support by HLD at HZDR, member of the European Magnetic Field Laboratory
(EMFL); by the Deutsche Forschungsgemeinschaft priority program SPP 1485,
by the German-Russian-Ukrainian research project by the Volkswagen-Stiftung;
and discussions with S.\ Johnston, D.\ Efremov, V.\ Grinenko, and
Y.\ Naidyuk are acknowledged.


\begin{thebibliography}{99}
\bibitem{rem_bc2}
The upper critical field of 8.25 T was accurately determined from the
torque (Fig.\ \ref{fig:torque}) and from specific-heat data (not shown).
\bibitem{rem_harm}
The additional non-sinusoidal feature appearing in the oscillating
sound-velocity signal above about 12 T is caused by a rather strong
higher-harmonic contribution in the $\alpha$-orbit oscillations, see
also Fig.\ 1 in Ref.\ \cite{ter95}.
\bibitem{rem_tau}
The rather evolved hysteretic torque signal in the mixed state is a
consequence of the torque-measurement technique. Indeed, the torque
is given by $\vec{\tau} = \vec{M}\times \vec{B}$, which means that
only the magnetization ($\vec{M}$) component perpendicular to
$\vec{B}$ is detected. Since even the parallel component of $\vec{M}$
depends strongly on many details (pinning centers, sweep rate, sample
shape), the perpendicular component appears to be unpredictable and
leads to the shown complicated field dependence.
\bibitem{rem_fund}
Here and in the following we only consider the contribution to the
fundamental frequency.
\bibitem{rem_mc}
Strictly speaking, $m_c$ extracted from the temperature-dependent damping
is renormalized by many-body interactions, whereas the mass in Eq.\ (\ref{damp})
is not (see \cite{jan98}). This point is usually neglected as we do as well
here.
\bibitem{rem_hexa}
Any deviation from an ideal hexagonal flux-line lattice would lead to
a stronger damping of the oscillating signal in the mixed state, as
seen in the torque data.
\bibitem{rem_gap}
Even when unphysically neglecting $R_{FL}$, i.e., assuming a much larger
$\kappa$, and using a lower bound for $T_D$, $\Delta_0$ would always be
less than 1 meV using Eq.\ (\ref{maki}).
\bibitem{rem_plot}
Note that this plot is valid only above $B_{c2}$ estimated using
Eq.\ (\ref{bc2alpha}) when the residual superconductivity (here in the
$\alpha$ pocket) is induced by interband coupling.
\end{thebibliography}
\end{document}